\documentclass[11pt,a4paper]{article}
\usepackage[T1]{fontenc}
\usepackage[utf8]{inputenc}
\usepackage{graphicx}
\usepackage{geometry}
\usepackage{amsmath}
\usepackage{amssymb}
\usepackage{tensor}
\usepackage{hyperref}

\newcommand{\hs}{\hspace{0.75pt}}

\title{Scalar Gravity and the Double Copy in the\\ Framework of General Relativity}
\author{Emanuelis Lazauskas$^a$\\[2pt]
\normalsize $^a$\textit{Center for Physical Sciences and Technology},\\[-2pt] \normalsize \textit{Savanorių Ave. 231, Vilnius, Lithuania}}
\date{}

\begin{document}

\maketitle

\begin{abstract}
    \noindent
    The classical double copy relations establish correspondence between exact gravity solutions and their counterparts in biadjoint and gauge theories. In this work, we investigate a peculiar case of double copy relations arising in exact solutions of the scalar gravity equation, relating static spacetime geometry to the Komar density distribution. We use a scalar-restricted static metric ansatz to isolate the degree of freedom governed by the linear equation for the zeroth copy. The corresponding metric solutions are interpreted as describing gravity without self-interaction, based on the Komar density distribution. The single copy of these metrics reveals direct correspondence between the sources of scalar gravity and electrostatics. Furthermore, this correspondence persists even at the level of particles, their motion being governed by the same non-relativistic expressions with respect to the background space. These and other results lead us to conclude that scalar gravity is the manifestation of double copy relations encoded in the structure of general relativity.
\end{abstract}

\section{Introduction}

Recently, there has been a lot of work devoted to studying relations between the gravity and gauge theories. The amplitudes program in quantum field theory has revealed one such relationship, known as the BCJ (Bern, Carrasco, Johansson) double copy [1–3], which at tree-level is equivalent to the well-known Kawai, Lewellen, Tye relations [4–10]. The original version of this theory came to be known as the double copy, since it relates the n-point graviton amplitude with two copies in Yang–Mills theory, facilitated by the interchange of color and kinematic factors. This relies on the gauge amplitude being expressible in the BCJ form, for which the color–kinematics duality is made manifest. The reverse procedure, i.e., the replacement of kinematic factors in the Yang–Mills amplitude with color factors, produces the zeroth copy, which describes amplitudes of the biadjoint scalar theory. Although the latter theory is not found in nature, by the theory of double copy it underlies the structure of the perturbative gauge and gravity theories. This is especially true since the BCJ double copy is believed to hold at loop level [11–29].

Given the success of the double copy prescription for amplitudes, steps were undertaken to explore similar relations at the classical level. Several studies demonstrated BCJ-like relationships existing between the classical solutions utilizing various perturbative approaches [20–24]. Of particular relevance is then a question as to what extent might an exact relation arise involving non-perturbative solutions. Some serious progress in this direction was achieved in [25], wherein the authors found an infinite class of exact solutions that matched under the double copy (we refer to this as the classical double copy). Specifically, they showed that time-independent Kerr–Schild metrics can be interpreted as double copies of solutions to linearized Yang–Mills equations. Besides the stationary Kerr–Schild spacetimes, identical relationships were subsequently extended to more general solutions, including time-dependent cases [26–31]. While no exact mapping between the classical version and the BCJ double copy has been established so far, a number of studies suggest that they are indeed related [21–28].

In this work, we consider a peculiar case of double copy relations arising in scalar-restricted Kerr–Schild metrics\footnote{These metrics are further assumed to be asymptotically flat and non-degenerate.}, minimally modified to achieve a static configuration. More precisely, spacetimes that are reconstructed from scalar characterized Kerr–Schild ansatz using a totally geodesic hypersurface and the associated unit normal vector by declaring the time-like killing field to be irrotational. The modified geometry is of interest to us since it admits a linearized Ricci tensor projection. A further motivation is the fact that this expression can be related to the Komar density distribution, thereby forming a self-consistent linear equation for the metric tensor. We therefore interpret the resulting solutions as describing gravity without self-interaction, as generated by the Komar density distribution. We refer to them as solutions of scalar gravity.

The comparison of these spacetimes with the gauge theory using double copy procedure reveals striking similarities on multiple levels. First, the single copy, stripped of its color factors, corresponds to an electrostatic potential produced by an arbitrary charge distribution. In fact, the double copy establishes direct correspondence between the solutions of scalar gravity and those of Maxwell’s equations. Second, the correspondence persists even between equations of motion involving particles. Specifically, these equations reduce to the same non-relativistic expressions provided that both theories evaluate them with respect to the background space of linear field equations. These results come with a number of intriguing implications, which will be examined through the course of this work. They lead us to interpret scalar gravity as a manifestation of double copy relations encoded in the framework of general relativity.

The content is organized as follows. In section 2, we briefly review the classical double copy of stationary Kerr–Schild metrics [25]. In section 3, the scalar-restricted static metric ansatz is derived, which serves as the constraint for scalar gravitational fields. Section 4 investigates the corresponding linearization of the Ricci tensor projection. In section 5, we compare the derived solutions with those of gauge theory using the double copy procedure. In the last section, a short summary is presented together with a few remarks.

\section{The classical double copy}

In this section, we present a brief review of the classical double copy [25], which relates time-independent Kerr–Schild spacetimes to solutions of the biadjoint and gauge theories. One may start with the assumption that the background is flat, and introduce a general Kerr–Schild ansatz
\begin{equation}
    \tensor{g}{_\mu_\nu}=\tensor{\eta}{_\mu_\nu}+\phi\hs\tensor{k}{_\mu}\tensor{k}{_\nu},
\end{equation}
where $\tensor{\eta}{_\mu_\nu}$ is the Minkowski metric, $\phi$ is a scalar function, and $\tensor{k}{_\mu}$ is a null vector with respect to both metrics $\tensor{g}{_\mu_\nu}$ and $\tensor{\eta}{_\mu_\nu}$, so that we have
\begin{equation}
    \tensor{g}{_\mu_\nu}\tensor{k}{^\mu}\tensor{k}{^\nu}=\tensor{\eta}{_\mu_\nu}\tensor{k}{^\mu}\tensor{k}{^\nu}=0\hspace{4pt}\text{and}\hspace{4pt}\tensor{g}{^\mu^\nu}=\tensor{\eta}{^\mu^\nu}-\phi\hs\tensor{k}{^\mu}\tensor{k}{^\nu}.
\end{equation}
Consequently, we see that the index of $\tensor{k}{^\mu}$ can be raised by either the Minkowski or the full metric. If we further suppose that $\tensor{k}{^\mu}$ is geodetic, i.e.,
\begin{equation}
    \tensor{k}{^\mu}\tensor{\nabla}{_{\!\mu}}\tensor{k}{^\nu}=\tensor{k}{^\mu}\tensor{\partial}{_\mu}\tensor{k}{^\nu}=0,
\end{equation}
the Ricci tensor can be expressed in a manifestly linear form:
\begin{equation}
    \tensor{R}{^\mu_\nu}=\frac{1}{2}\big(\tensor{\partial}{^\mu}\tensor{\partial}{_\alpha}\left(\phi\hs\tensor{k}{^\alpha}\tensor{k}{_\nu}\right) +\tensor{\partial}{_\nu}\tensor{\partial}{_\alpha}(\phi\hs\tensor{k}{^\alpha}\tensor{k}{^\mu})-\partial^2(\phi\hs\tensor{k}{^\mu}\tensor{k}{_\nu})\big).
\end{equation}
Additional simplification occurs when consideration is restricted to stationary metrics (with $\tensor{\partial}{_0}\phi=\tensor{\partial}{_0}\tensor{k}{_\mu}=0$) that have a Kerr–Schild vector $\tensor{k}{^\mu}=(1,\hat{k})$. In such a configuration, the Ricci tensor components reduce to
\begin{subequations}
    \renewcommand{\theequation}{\theparentequation.\arabic{equation}}
    \begin{alignat}{2}
    &\tensor{R}{^\mu_0} &{}={}& -\frac{1}{2}\tensor{\partial}{_\nu}\big(\tensor{\partial}{^\mu}(\phi\hs\tensor{k}{^\nu})-\tensor{\partial}{^\nu}(\phi\hs\tensor{k}{^\mu})\big) , \\
    &\tensor{R}{^i_j} &{}={}& \frac{1}{2}\tensor{\partial}{_l}\big(\tensor{\partial}{^i}(\phi\hs\tensor{k}{^l}\tensor{k}{_j})+\tensor{\partial}{_j}(\phi\hs\tensor{k}{^l}\tensor{k}{^i})-\tensor{\partial}{^l}(\tensor{k}{^i}\tensor{k}{_j})\big) .
    \end{alignat}
\end{subequations}
In the case of the Einstein vacuum equation $\tensor{R}{_\mu_\nu}=0$, these expressions establish direct correspondence between exact solutions of the gravity, gauge, and biadjoint theories.

Specifically, starting from the Kerr–Schild metric Eq. (1) (referred to as the double copy), one can construct a non-abelian gauge field (called the single copy) via
\begin{equation}
    A^a_\mu=c^a\phi\hs\tensor{k}{_\mu},
\end{equation}
where $c^a$ are arbitrary constant color factors [31]. If the double copy satisfies the Einstein equation, the single copy is guaranteed to satisfy linearized Yang–Mills equations, provided that the couplings and sources are suitably replaced. In this work, we focus on the abelian character of the equations, and hence consider the field $\tensor{A}{^\mu}=\phi\hs\tensor{k}{^\mu}$ without the color factors. This allows us to identify $\tensor{A}{^\mu}$ with electromagnetic (E\&M) vector potential, sourced by the current [28]
\begin{equation}
    \tensor{j}{^\mu}=-\tensor{R}{^\mu_0}/2\pi.
\end{equation}
The same procedure of replacing the Kerr–Schild vector with color factors can be repeated to obtain the biadjoint scalar (the zeroth copy)
\begin{equation}
    \phi^{aa'}=c^a \tilde{c}^{\hspace{0.5pt}a'}\phi
\end{equation}
featuring an additional set of factors $\tilde{c}^{\hspace{0.5pt}a'}$, which may be associated with a different gauge group than that of $c^a$. This field then satisfies the linearized equations for the biadjoint scalar field
\begin{equation}
    \Delta\phi^{aa'}=c^a \tilde{c}^{\hspace{0.5pt}a'}\Delta\phi=0.
\end{equation}
Although the $\phi^{aa'}$ is not directly applicable to nature, it serves as the building block of amplitude numerators in gauge and (correspondingly) gravity theories [32]. In the classical context, it provides one argument that the double copy considered here is indeed related to the BCJ story for amplitudes [25]. This is the same as before, for our purposes we consider the field $\phi$ stripped of its color indices.

As a final note, we draw attention to the fact that we are dealing with linear field equations. It is hence possible to characterize the gravitational field in terms of E\&M vector potentials. The classical double copy procedure describes exactly how to do it: by taking the single copy of the Kerr–Schild double copy. This formalism plays a crucial role in our derivation of scalar gravitational fields, as well as their comparison with electrostatics.

\section{Scalar gravity ansatz}
As was remarked in the previous section, the classical double copy relations applied to linear field equations enable characterization of the gravitational field in terms of E\&M vector potentials. This allows us to perceive scalar gravitational fields as embodied by those solutions that feature single copies corresponding to electrostatic potentials. As an example, consider the static spherically symmetric Kerr–Schild metrics discussed in [28],
\begin{equation}
    \tensor{g}{_\mu_\nu^{\hspace{-10pt}\scalebox{0.65}{KS}}}=\tensor{\eta}{_\mu_\nu}+\phi(r)\tensor{k}{_\mu}\tensor{k}{_\nu},
\end{equation}
where $\tensor{k}{^\mu}=(1,\hat{r})$. The single copy of these metrics is
\begin{equation}
    \tensor{A}{^\mu}=\phi\hs(1,\tensor{k}{^i}).
\end{equation}
As the result of spherical symmetry, the “magnetic potential” $\tensor{A}{^i}$ can be gauged away by a transformation $\tensor{A}{^\mu}\rightarrow\tensor{A}{^\mu}+\tensor{\partial}{^\mu}\lambda$, where
\begin{equation}
    \lambda=-\int^{r}\!\phi(r')\,dr',
\end{equation}
leaving only the scalar potential $\tensor{A}{^\mu}=(\phi,\vec{0}\hs)$. Hence, the spherically symmetric Kerr–Schild solutions provide a prime example of scalar gravitational fields, the single copies of which correspond to the electrostatic potential of a spherically symmetric charge distribution $\tensor{j}{^\mu}=(\rho(r),\vec{0}\hs)$.

In this work, we shall demonstrate that there exists a sensible generalization of scalar gravitational fields for the arbitrary symmetry. More specifically, that it is possible to construct meaningful metrics, which in the case of the single copy correspond to the electrostatic potential of an arbitrary charge distribution $\rho(x^i)$. Toward this end, we first generalize metrics for the ansatz Eq. (10) in order to isolate the degree of freedom responsible for scalar gravity. Accordingly, let us introduce the scalar-restricted Kerr–Schild ansatz,
\begin{equation}
    \tensor{g}{_\mu_\nu^{\hspace{-10pt}\scalebox{0.66}{KS}}}=\tensor{\eta}{_\mu_\nu}+\phi(x^i)\tensor{k}{_\mu}\tensor{k}{_\nu},
\end{equation}
where the 3-vector $\tensor{k}{^i}$ (with $\tensor{k}{^0}=1$) is defined as a normalized gradient of $\phi$ (with respect to $\tensor{\eta}{_i_j}$),
\begin{equation}
    \tensor{k}{^i}=(\nabla\phi)^i/\sqrt{(\nabla\phi)^2}.
\end{equation}
In order to achieve a static configuration, the geometry of metrics Eq. (13) requires the elimination of gravitomagnetic currents. This can be done by reconstructing spacetimes from a totally geodesic hypersurface $\Sigma$ and associated unit normal vector $\tensor{n}{^\mu}$ by declaring the time-like killing field $\tensor{\partial}{_0}$ (parallel to $\tensor{n}{^\mu}$) to be irrotational. Such alteration of Kerr–Schild geometry is performed throughout the remainder of this section.

For this purpose, let us invoke some elements of the embedding formalism. Particularly, let $\psi:\Sigma\rightarrow M$ be an embedding map of a totally geodesic hypersurface $\Sigma$  (with vanishing extrinsic curvature $\tensor{K}{_a_b}$) in a manifold $M$ (equipped with the metric $\tensor{g}{_\alpha_\beta^{\hspace{-10.5pt}\scalebox{0.65}{KS}}}\hs$). Given a curvilinear coordinate system $(t,\xi^a)$ on $M$ and $(\xi^a)$ on $\Sigma$, the push-forward $\psi_*:T_p\Sigma\rightarrow T_p M$ defines a mapping of spatial basis vectors $\tensor{\partial}{_{\bar{a}}}$ in $T_p\Sigma$  to their extensions $\tensor{\partial}{_{\bar{\alpha}}}$ in $T_p M$ ($\tensor{\partial}{_\alpha}$ projected onto $T_p M$). Conversely, the pull-back $\psi^*:T^*_p M\rightarrow T^*_p \Sigma$ induces basis one-forms $\text{d}\tensor{\bar{\xi}}{^a}$ in $T^*_p M$ ($\text{d}\xi^\alpha$ restricted to $\Sigma$) from basis one-forms $\text{d}\xi^a$ in $T^*_p M$ (satisfying $\langle \text{d}\tensor{\bar{\xi}}{^a},\tensor{\partial}{_{\hspace{0.25pt}\bar{b}}}\rangle=\tensor{\delta}{^{\bar{a}}_{\bar{b}}}$). It also induces a metric $\tensor{\gamma}{_a_b}$ on the hypersurface $\Sigma$. However, the embedding $\psi$ provides these mappings only in one direction, it does not allow to carry them in reverse.

\noindent It is therefore convenient to define an orthogonal projector $\vec{\gamma}:T_p M\rightarrow T_p \Sigma$,
\begin{equation}
    \tensor{\gamma}{^\alpha_\beta}=\tensor{\delta}{^\alpha_\beta}+\tensor{n}{^\alpha}\tensor{n}{_\beta},
\end{equation}
which provides the reverse mappings. The normal vector of the hypersurface $\Sigma$  is chosen to be parallel to the killing field $\tensor{\partial}{_0}$ and with the condition $n^2=-1$, having the components
\begin{equation}
    \tensor{n}{^\alpha}=\big(1/\sqrt{1-\phi},\vec{0}\hspace{1pt}\big).
\end{equation}
This specifies hypersurfaces that are totally geodesic; i.e., they have a vanishing extrinsic curvature
\begin{equation}
    \tensor{K}{_\alpha_\beta}=-\frac{1}{2}\tensor{\gamma}{^\gamma_\alpha}\tensor{\gamma}{^\delta_\beta}\tensor{\mathcal{L}}{_n}\hs \tensor{g}{_\gamma_\delta^{\hspace{-9pt}\scalebox{0.65}{KS}}}=0.
\end{equation}
With the hypersurfaces specified, the extension of $\tensor{\gamma}{_a_b}$ can be conveniently expressed in terms of the extended induced coordinate basis $\text{d}\tensor{\bar{\xi}}{^\alpha}$ as
\begin{equation}
    \tensor{\gamma}{_\alpha_\beta}=\tensor{g}{_\alpha_\beta^{\hspace{-10.5pt}\scalebox{0.65}{KS}}}\,+\tensor{n}{_\alpha}\tensor{n}{_\beta}.
\end{equation}
The resulting line element is given by
\begin{equation}
    ds^2=\frac{\phi}{1-\phi}\,\tensor{k}{_i}\tensor{k}{_j}\text{d}\tensor{\bar{\xi}}{^{\hs i}}\text{d}\tensor{\bar{\xi}}{^{\hspace{0.5pt} j}}.
\end{equation}
At this point we introduce a modification of the original Kerr–Schild geometry by declaring that the time-like killing field $K=\tensor{\partial}{_0}$ is irrotational,
\begin{equation}
    \tensor{\nabla}{_{\!\alpha}}\tensor{K}{_\beta}=\tensor{\nabla}{_{\!\beta}}\tensor{K}{_\alpha},
\end{equation}
where $\tensor{\nabla}{_{\!\alpha}}$ is the covariant derivative associated with the new connection. The condition Eq. (20) together with specified data on $\Sigma$ (induced metric $\tensor{\gamma}{_a_b}$ and extrinsic curvature $\tensor{K}{_a_b}$) and the original components of the unit normal vector $\tensor{n}{^\alpha}$ fixes the geometry of new spacetimes $(N,\tensor{g}{_\alpha_\beta})$. The full line element can be explicitly written as
\begin{equation}
    ds^2=-(1-\phi)\hs\text{d}t^2+\frac{\phi}{1-\phi}\bigg(\frac{\tensor{\partial}{_i}\phi\,\tensor{\partial}{_j}\phi}{\tensor{\eta}{^k^l}\tensor{\partial}{_k}\phi\,\tensor{\partial}{_l}\phi}\bigg)\text{d}\xi^{i}\text{d}\xi^{j},
\end{equation}
where the bar over spatial coordinates $\xi^i$ was removed to indicate that in the static geometry, basis vectors $\tensor{\partial}{_i}$ are tangent to slices of the foliation $(\Sigma_t)_{t\in\mathbb{R}}$. These metrics are assumed to be asymptotically flat as well as non-degenerate\footnote{These conditions are assumed to hold for the rest of the work, we shall not repeat them.}. Given these constraints, we refer to Eq. (21) as the ansatz for scalar gravity.

The reconstructed spacetimes given by Eq. (21) represent a minimally altered Kerr–Schild geometry of Eq. (13) that has a static configuration. As the result of using a totally geodesic hypersurface for slices of the foliation, the associated tangent geodesics in the new spacetimes are left unchanged. Additionally, the same remain and 4-acceleration vectors of static observers $u^\alpha=n^\alpha$ (orthogonal to time-symmetric $\Sigma$ and $\Sigma_t$), accompanied only by locally transformed time coordinate resulting in $u^0=0$ for the case of static metrics. As was already mentioned, the derived ansatz is of interest since it constrains the degree of freedom governing scalar gravitational fields. The related details are investigated in the next section.

\section{Scalar gravity}

We now restrict our attention to static spacetimes equipped with the metrics from Eq. (21) and show that they admit a linearized Ricci tensor projection perpendicular to the discussed slices $\Sigma_t$ of the foliation. This endeavor is considerably simplified by the fact that contraction with $n^\mu$ (Eq. 16) leaves only the $\tensor{R}{_0_0}$ component to be calculated
\begin{equation}
    \tensor{R}{_\mu_\nu}\tensor{n}{^\mu}\tensor{n}{^\nu}=N^{-2}\tensor{R}{_0_0},
\end{equation}
where the introduced convenience function $N=\sqrt{1-\phi}$ has a physical interpretation of relating coordinate time $t$ with the proper time $\tau$ measured by the static observer. In order to proceed, let us choose for simplicity the Cartesian coordinates $(\xi^i)=(x,y,z)$. With these changes the line element Eq. (21) reads
\begin{equation}
    ds^2=-N^2\text{d}t^2+\phi\hspace{0.5pt} N^{-2}\bigg(\frac{\tensor{\partial}{_i}\phi\,\tensor{\partial}{_j}\phi}{\tensor{\eta}{^k^l}\tensor{\partial}{_k}\phi\,\tensor{\partial}{_l}\phi}\bigg)\text{d}\xi^{i}\text{d}\xi^{j}.
\end{equation}
Since condition $\sqrt{-g}=1$ is satisfied, the Ricci tensor components can be calculated using a simplified formula:
\begin{equation}
    \tensor{R}{_0_0}=\tensor{\partial}{_\alpha}\tensor{\Gamma}{^\alpha_0_0}-\tensor{\Gamma}{^\alpha_0_\beta}\tensor{\Gamma}{^\beta_0_\alpha}.
\end{equation}
Equation (24) involves only a portion of Christoffel symbols $\tensor{\Gamma}{^\alpha_0_\beta}$. After some manipulation, they can be shown to have simple expressions:
\begin{subequations}
    \renewcommand{\theequation}{\theparentequation.\arabic{equation}}
    \begin{alignat}{2}
    &\tensor{\Gamma}{^0_0_0} &{}={}& 0, \\
    &\tensor{\Gamma}{^i_0_0} &{}={}& -N^2/2\cdot\tensor{\partial}{_i}\phi, \\
    &\tensor{\Gamma}{^0_0_j} &{}={}& -N^{-2}/2\cdot\tensor{\partial}{_j}\phi, \\
    &\tensor{\Gamma}{^i_0_j} &{}={}& 0.
    \end{alignat}
\end{subequations}
Substituting these expressions into Eq. (24) and then into Eq. (22) results in the major simplification
\begin{equation}
    \tensor{R}{_\mu_\nu}=-\frac{\nabla^2\phi}{2}.
\end{equation}
The above expression is equivalent to the linearized $\tensor{R}{^0_0}$ component given in Eq. (5) using Kerr–Schild coordinates. We have thus demonstrated the linearization of the Ricci tensor projection that occurs with respect to the metric ansatz Eq. (21).

As the result, these metrics possess a particularly tractable Komar energy integral. Given the killing vector discussed earlier, $K=\tensor{\partial}{_0}$, we can calculate the Komar energy contained within a 3-volume $V$ using
\begin{equation}
    E=\frac{1}{4\pi}\int_{V}\!\tensor{R}{_\mu_\nu}\tensor{n}{^\mu}\tensor{K}{^\nu}\sqrt{\gamma}\,d^3\xi ,
\end{equation}
where $\gamma=N^{-2}$ is the determinant of the induced metric $\tensor{\gamma}{_a_b}$. Exploiting expression Eq. (27), the integral evaluates to
\begin{equation}
    E=-\frac{1}{8\pi}\int_{V}\!\nabla^2\phi\,d^3\xi .
\end{equation}
The above expression shows that the Komar energy integral Eq. (27) reduces to Gauss’s law. In differential form, it is recovered as Poisson’s equation, featuring Komar density as the source,
\begin{equation}
    \rho_K=-\nabla^2\phi/8\pi.
\end{equation}
Consequently, we can think of solutions to this equation as the zeroth copies $\phi$ of the double copy Eq. (21), which then satisfies the gravity equation
\begin{equation}
    \tensor{R}{_\mu_\nu}\tensor{n}{^\mu}\tensor{n}{^\nu}=-4\pi\rho_K.
\end{equation}

Although Eq. (30) is not guaranteed to parameterize valid initial data, it relates spacetime geometry to Komar density distribution, and hence is self-consistent (albeit underdetermined). From Eq. (27), we can see that it simply assumes the fixed form of determinant $\gamma$, so as to satisfy $\sqrt{-g}=1$ in the Cartesian coordinate system. The imposition of the metric ansatz Eq. (21) then restricts the remaining degrees of freedom leaving only one. The fact that the latter is then linearized allows us to interpret the double copy as describing gravity without self-interaction, as generated by the Komar density $\rho_K$. Based on these considerations, we refer to the metrics of Eq. (21) satisfying Eq. (30) as solutions of scalar gravity. In the next section, it will be shown that this type of gravity is compliant with the definition of scalar gravitational fields based on electrostatic E\&M vector potentials.

This leaves one question to be answered: how to interpret the linearization of the Ricci tensor projection that occurs with respect to the scalar gravity ansatz? The Kerr–Schild metrics have a very clear interpretation in this regard, since linearization of the expressions in Eq. (5) can be associated with the vanishing of the pseudo-energy–momentum tensor [33]. However, this phenomenon occurs only in very rare cases, those in which spacetimes are algebraically special [34]. This naturally excludes all the metrics considered here (apart from spherically symmetric solutions). Instead, we focus on the observation that the only degree of freedom of Eq. (30) constrained by scalar gravity ansatz is governed solely by the equation for the zeroth copy. This leads us to the hypothesis that the corresponding linearization is related to the double copy relations encoded in the framework of general relativity. In contrast to the Kerr–Schild double copy, which relies on linearization, we claim that the double copy considered here is the cause of it.

This statement comes with a share of intriguing implications. If the discussed interpretation is correct, it would provide further evidence for the double copy being a fundamental relation between different types of theories. It would also suggest that a similar linearization may be possible for the single copy involving axially symmetric stationary spacetimes, as they happen to admit the Komar surface integral for the angular momentum. Presumably, a non-trivial linearization of perpendicular and mixed Ricci tensor projections (with respect to $\Sigma_t$) that could be related to the Komar energy and angular momentum densities. This would then define a self-consistent system of “vector gravity” equations, solutions to which would single copy to the general class of stationary axisymmetric E\&M solutions. While at the present these are just speculations, one path to the truth lies in an understanding of the role the metric ansatz Eq. (21) plays in the linearization process. We shall return to this question in the next section.

\section{Comparison with gauge theory}

In the previous section, it was shown that scalar gravity solutions can be interpreted as describing gravity without self-interaction, as generated by the Komar density distribution. In this section, we compare them with the gauge theory using the double copy procedure. In order to do that, we first address the ambiguity associated with taking the single copy of metric ansatz Eq. (21). The standard procedure outlined in [25] cannot be applied directly, since in our case the (non-perturbative) graviton does not decompose into the product of two Lorentz vectors. However, it can be applied to the stationary Kerr–Schild ansatz Eq. (13), resulting in the E\&M vector potential
\begin{equation}
    A^\mu=\phi \big(1,(\nabla\phi)^i)/\sqrt{(\nabla\phi)^2}\big) .
\end{equation}
This vector potential needs to be further modified to account for the elimination of gravitomagnetic currents. Since on the gravity side this action represents the least possible modification of Kerr–Schild geometry to achieve a static configuration (see section 3), on the E\&M side we can identify it via the elimination of magnetic currents. The resulting potential (in the Coulomb gauge) and its field strength are
\begin{subequations}
    \renewcommand{\theequation}{\theparentequation.\arabic{equation}}
    \begin{alignat}{2}
    &A^\mu &{}={}& (\phi,\vec{0}\hs), \\
    &\tensor{F}{^\mu_\nu} &{}={}& \partial^\mu A^\nu-\partial^\nu A^\mu.
    \end{alignat}
\end{subequations}
By the Gauss’s law Eq. (28), the field $A^\mu$ is guaranteed to satisfy vacuum Maxwell’s equations, while associating the Komar density $\rho_K$ with charge density $\rho$ allows us to match even their sources
\begin{equation}
    \tensor{\partial}{_\nu}\tensor{F}{^\mu^\nu}=4\pi \tensor{j}{^\mu} ,
\end{equation}
where $j^\mu$ is calculated according to Eq. (7) (see section 2). Hence, the single copy is just an electrostatic potential produced by the charge distribution $j^\mu=(\rho,\vec{0}\hs)$, complying with the definition of scalar gravitational fields given in section 3. The double copy has established direct correspondence between the sources of scalar gravity and those of gauge theory.

After exploring the relations between the equations of motion for fields, we now turn to investigate the equations of motion for particles. In general, one would not expect similar relations to occur in this context, because relativistic effects in curved spacetimes are drastically different from those in a flat space. Nevertheless, it is possible to ignore these effects by restricting consideration to static particles. It can then be shown that the acceleration field of static observers $a_G\in T_p U$ corresponds to the electrostatic acceleration of charged particles $a_{EM}\in T_{\xi(p)}V$, and through isomorphism
\begin{equation}
    \xi_{*}:T_p U\rightarrow T_{\xi(p)}V ,
\end{equation}
where $U\subset M$ and $V\subset \mathbb{R}^4$ are respective subsets and the chart $(U,\xi)$ is assumed to be a local diffeomorphism, i.e., the composition $\xi\circ\xi^{-1}:\mathbb{R}^4\rightarrow\mathbb{R}^4$ is well-defined and smooth. The statement can be demonstrated by considering a push-forward mapping of the coordinate basis
\begin{equation}
    \xi_{*}\frac{\partial}{\partial \xi^{\mu}}\bigg|_p=\frac{\partial}{\partial \xi^\mu}\bigg|_{\xi(p)},
\end{equation}
which when using Cartesian coordinates implies a coordinate basis in $T_p U$ being mapped to the orthonormal basis in $T_{\xi(p)}V$. In this basis, the 4-acceleration of a static observer can be evaluated as
\begin{equation}
    \tensor{n}{^\alpha}\tensor{\nabla}{_\alpha}\tensor{n}{^\mu}=-\frac{(\nabla\phi)^i}{2},
\end{equation}
which can be interpreted as acceleration of static particles ($u^0=1$) with equal mass and charge ($m=q$) situated in an electrostatic field,
\begin{equation}
    \frac{d u^\mu}{d\tau}=\frac{q}{m}\tensor{F}{^\mu^\alpha}\tensor{u}{_\alpha}=\frac{(\nabla\phi)^i}{2} .
\end{equation}
Consequently, as the result of dealing with a static geometry in which the effects of gravity are independent of velocity, the same 3-acceleration applies to any observer that is locally inertial with respect to the Minkowski metric. Hence, we discover that the equations of motion for particles have the same non-relativistic form provided that both theories evaluate them with respect to the Minkowski metric. We have shown that similar to the double copy relations for fields, there exist comparable relations for particles.

This raises the question of whether there is a deeper connection between these notions of motion involving fields and particles that is ensuring correspondence with the gauge theory on both sides. Restating this based on mathematical facts leads to the following. One the one side, we have a gravity equation Eq. (30) that is linear in the zeroth copy. On the other side, the geodesic equation is also linear in the zeroth copy. Both of these manifest with respect to the background space. Instead of treating this as a mere coincidence, we speculate it to be the characteristic of gravity without self-interaction. It then suggests that one way of validating the concept of “vector gravity” is by seeking a metric ansatz with similar linear geodesic equations for the stationary axisymmetric configuration. Evidently, this endeavor is more challenging, since in addition to “electric” gravity, one is required to account for gravitomagnetic effects. Nevertheless, it could reveal undiscovered insights into the double copy theory as well as reconfirm or discard suspicions presented in the previous section.

\section{Concluding remarks}

We have investigated the special case of double copy relations arising in the framework of general relativity, specifically, the case involving solutions of the scalar gravity equation, relating static spacetime geometry with the Komar density distribution. We first derived the scalar gravity ansatz, which isolates the degree of freedom governed by the linear equation for the zeroth copy. The corresponding solutions were interpreted as describing gravity without self-interaction, based on the Komar density distribution. We then compared them with gauge theory using the double copy prescription. It revealed a direct correspondence between sources of scalar gravity and electrostatics. We found that this correspondence persists even at the level of particles, their motion being governed by the same non-relativistic expressions with respect to the background space. Instead of treating this as a mere coincidence, we considered it to be characteristic of gravity without self-interaction. These and other results led us to the conclusion that scalar gravity is the manifestation of double copy relations encoded in the structure of the general theory of relativity.

This statement comes with a share of intriguing implications. If the discussed interpretation is correct, it would provide further evidence for the double copy being a fundamental relation between the different types of theories. It would also suggest that a similar linearization may be possible for the single copy involving axially symmetric stationary spacetimes, as they happen to admit the Komar surface integral for the angular momentum. Presumably, a non-trivial linearization of Ricci tensor projections could be related to the Komar energy and angular momentum densities. This would then define a self-consistent system of “vector gravity” equations, solutions to which would single copy to a general class of stationary axisymmetric E\&M solutions. While at the present these are just speculations, a possible way to confirm them is by seeking a metric ansatz with analogous geodesic characteristics for the stationary axisymmetric configuration. This presents one prospective direction in which this work could be continued to potentially reveal undiscovered insights into the nature of classical double copy relations.

\section*{Note from the author}

This work is intended for drawing attention to the novel double copy relations arising in the framework of general relativity. The known issue pertaining the current version of the article is the need for a more careful and detailed analysis of results.

\end{document}